\documentclass[prl,twocolumn,superscriptaddress,showpacs]{revtex4}

\usepackage{mathbbol,amsmath,amsfonts,amssymb,bm,color}
\usepackage{graphicx,psfrag}

\newcommand{\half}{\mbox{$\textstyle \frac{1}{2}$} }

\newcommand{\hh}{\begin{picture}(13,9)(-2,2)
    \put (0,0) {\line (1,0) {8}}
    \put (8,8) {\line (-1,0) {8}}
    \put (0,0) {\circle*{3}}
    \put (0,8) {\circle*{3}}
    \put (8,0) {\circle*{3}}
    \put (8,8) {\circle*{3}}
    \end{picture}
}
\newcommand{\vv}{\begin{picture}(13,9)(-2,2)
    \put (0,0) {\line (0,1) {8}}
    \put (8,8) {\line (0,-1) {8}}
    \put (0,0) {\circle*{3}}
    \put (0,8) {\circle*{3}}
    \put (8,0) {\circle*{3}}
    \put (8,8) {\circle*{3}}
    \end{picture}
}

\begin{document}

\title{Unified approach to Quantum and Classical Dualities}

\author{E. Cobanera}
\affiliation{Department of Physics, Indiana University, Bloomington,
IN 47405, USA}
\author{G. Ortiz}
\affiliation{Department of Physics, Indiana University, Bloomington,
IN 47405, USA}
\author{Z. Nussinov}
\affiliation{Department of Physics, Washington University, St.
Louis, MO 63160, USA}

\date{\today}

\begin{abstract}
We show how classical and quantum dualities, as well as duality
relations that appear only in a  sector of certain theories ({\em
emergent dualities}), can be unveiled, and systematically established.
Our method relies on the use of  morphisms of the {\em bond algebra} of
a quantum Hamiltonian. Dualities are  characterized as unitary mappings
implementing such morphisms, whose even powers become symmetries of the
quantum problem. Dual variables
-which were guessed in the past- can be derived in our formalism. We
obtain new self-dualities for four-dimensional Abelian gauge field
theories.
\end{abstract}
 \maketitle

{\it Introduction.} Dualities appear in nearly all disciplines of
physics and play a central role in statistical mechanics and field
theory \cite{Savit,witten}. When available, these mathematical
transformations provide an elegant, efficient way to obtain information
about models that need not be exactly solvable. Most notably, dualities
may be used to determine features of phase diagrams such as boundaries
between phases, and the exact location of some critical/multicritical
points. Historically, dualities were introduced in classical statistical
mechanics by Kramers and Wannier (KW) as a relation between the
partition function of one system at high temperature (or weak coupling)
to the partition function of another (dual) system at low temperatures
(or strong coupling). This relation allowed for a determination of the
exact critical temperature of the two-dimensional Ising model on a
square lattice \cite{KW}, before the exact solution of the model was
available. Later on, it was noticed that, due to the connection between 
quantum theories in $d$ space dimensions and classical statistical
systems in $d+1$ dimensions, dualities can provide relations between
quantum theories in the strong coupling and weak coupling regimes
\cite{Savit}. The current work is motivated by a quest for a simple
unifying framework for the detection and treatment of dualities.

We will describe an algebraic approach to dualities and self-dualities
for systems of {\it arbitrary spatial dimensionality {\cal d}}. We will
show that quantum (self-)dualities (a connection between Hamiltonians)
become dualities of the related classical statistical problem in $d+1$
dimensions. Thus, {\it quantum and classical (self-)dualities are
intrinsically equivalent}, yet it will become clear that quantum
(self-)dualities are -with the technique presented here- much easier to
detect and exploit. The gist of the method is the characterization of
quantum (self-)dualities as structure preserving mappings
(homomorphisms) between operator algebras {\it which are Hamiltonian
dependent}. The structure of quantum mechanics further requires that
these (self-)duality mappings should be unitarily implementable. In
contrast, generalized Jordan-Wigner transformations \cite{GJW} for
example, are dictionaries connecting representations, independent of the
structure of any particular Hamiltonian.

{\it Bond Algebras and Dualities.} Our main thesis is that quantum
dualities (self-dualities) are homomorphisms (automorphisms) of {\it
bond algebras} \cite{bondDec08} that preserve locality of interactions
and can be implemented through a unitary map. Take a quantum Hamiltonian
$H$, given as a sum of {\it quasi-local} operators or {\it bonds}
$\{h_{R}\}$ weighed by couplings $\alpha_R$, $H = \sum_{R} \alpha_R
h_{R}$. The index $R$ can represent, for example, lattice sites. The
{\it bond algebra} of $H$, ${\cal A}_H$, is the smallest operator
algebra that contains every bond in $H$ -and thus $H$ itself. It can be
described as the algebra of all linear combinations of  products of
bonds $\prod h_R$ and the identity operator. The core idea is that two
Hamiltonians $H$ and $H_{\sf dual}$ are dual to each other if there is a
unitarily implementable homomorphism $\Phi$ between their bond algebras
mapping $H$ to $H_{\sf dual}$ { \it up to irrelevant terms in the
thermodynamic limit} \cite{irrelevant}. So we demand that $\Phi(H)=
U_DHU_D^{\dagger} = H_{\sf dual}+V_{\sf B}$ where the boundary operator
$V_{\sf B}$ is irrelevant \cite{irrelevant}. If $H$ and $H_{\sf dual}$
share the same bonds but with {\it different} couplings, then the
duality is nothing but a self-duality,  established through an
automorphism of ${\cal A}_H$. This scenario includes the very useful
special case of two exchanged couplings representing a {\it {\bf weak
coupling}}$\leftrightarrow${\it {\bf strong coupling}} exchange. To make
clear that this approach is physically sensible, it is enough to notice
that such homomorphisms {\it preserve the Heisenberg equations of
motion}. Notice that the labels $\{R\}$ are completely arbitrary, no
reference is made to any particular geometry or dimensionality. The
primary algebraic objects are the bonds \cite{bondDec08}, built out of
elementary degrees of freedom such as spins. In the past, quantum
dualities such as KW were presented as non-local mappings between
elementary degrees of freedom. In contrast, duality morphisms are
mappings local in the bonds and, remarkably, provide means to derive
those non-local mappings (which shows that these self-duality
automorphisms are indeed the quantum version of the classical
order-disorder transformations of Kadanoff and Ceva \cite{KC}).  That
all dualities are manifestations of bond algebraic morphisms is not
obvious.
If, however, as is the standard case, two systems are dual to one
another on {\em general subsets} $\Lambda$ of an infinite lattice then
an exact duality between the two systems exists {\em if and only if} the
bond algebras of the two systems are identical. The proof of this
assertion is straightforward. The proviso of {\em general sublattices}
implies that a unitary transformation giving rise to the same spectrum
may be applied for a general collection of bonds $R \in \Lambda$ and
their duals $R^{\prime} \in \Lambda'$: $U_{D} H U_{D}^{\dagger} = U_{D}
\Big( \sum_{R \in \Lambda} \alpha_{R} h_{R} \Big) U_{D}^{\dagger} =
\sum_{R' \in \Lambda'} \alpha_{R'}  h_{R'}^{\prime} = H_{\sf dual}$. As
this holds for all $\Lambda$, it follows that $U_{D} h_{R}
U_{D}^{\dagger} = h^{\prime}_{R'}$ for {\em all} $R,R'$.  If two sets of
operators (including the bond operators $\{h_{R}\}$) are related by a
unitary transformation $U_{D}$ then their algebras are identical.
Similarly, if two sets of operators $\{h_{R} \}$ and
$\{h^{\prime}_{R'}\}$ exhibit an identical algebra then there is a
unitary transformation $U_{D}$ relating them.

In general, self-dualities do not leave $H$ invariant. {\it They are
symmetries of the bond algebra ${\cal A}_H$}, and this is the  key to
detect them. However, they may become symmetries on appropriate regions
of parameter space. If, e.g., $U_D$ exchanges the couplings $g$ and $g'$
in $H$ then at the self-dual point $g=g'$, $[H,U_D]=0$ (up to the
irrelevant terms \cite{irrelevant}). Moreover, if $U_D$ effects the
exchange for any values of $g$ and $g'$, then for even $n$,
$[U_D^{n},H]=0$  (again, up to irrelevant terms). Taking $n=2$ we see
that
$$\boxed{\mbox{\sf Self-duality}\rightarrow \sqrt{{\sf
Quantum\ Symmetry}}}.$$
{\it Thus a self-duality could reveal nontrivial hidden symmetries of a
problem}. Of course, the symmetries $U_D^{n}$, need not be all
independent or non-trivial  (we will see examples below). One can always
add to $H$ an irrelevant boundary term $V_{\sf B}'$ (related, but not
equal to $V_B$) derived from the bond algebra, so that even for finite
systems $[H+V_{\sf B}',U_D^n]=0$ {\it exactly}.  Thus, it may be useful
to work with the more symmetric $H+V_{\sf B}'$.

As a basic illustration, take  
$\tilde{H}[j,h]=j\sum_{l=1}^{N-1}\sigma^z_l\sigma^z_{l+1}+h\sum_{l=1}^N
\sigma_l^x+j\sigma^z_N=H[j,h]+j\sigma^z_N$, $(j\sigma^z_N=V_{\sf B}')$
(the $\sigma^\alpha_l$ are Pauli matrices), where $H[j,h]$ is the
Hamiltonian of an Ising chain in a transverse magnetic field ($N$
spins). One can check that $\sigma^x_1 \mapsto\sigma^z_N$, $\sigma^z_N
\mapsto\sigma^x_1$, $\sigma^x_i \mapsto\sigma^z_{r(i)}
\sigma^z_{r(i)+1}$ ($i=2,3,\cdots,N$), $\sigma^z_i \sigma^z_{i+1}\mapsto
\sigma^x_{r(i)}$ ($i=1,\cdots,N-1$), with $r(i)=N+1-i$, gives a
unitarily implementable  automorphism $\Phi$ of $H$'s bond algebra.
$\Phi$ is clearly a self-duality for the Ising chain $H[j,h]$, $U_D
H[j,h]U_D^{\dagger}=H[h,j]+V_{\sf B}$, with boundary term  $V_{\sf
B}=h\sigma^z_N-j\sigma^x_1$, and it is an {\it exact} self-duality for
$\tilde{H}$, $\Phi(\tilde{H}[j,h])=
U_D\tilde{H}[j,h]U_D^{\dagger}=\tilde{H}[h,j]$. In this simple case,
$U_D^2=1$. The standard approach \cite{kogut} to this self-duality
involves defining non-local spin operators -the dual variables- but
nothing in principle determines their form; dual variables have to be
guessed. In contrast, in our formalism it is natural to use the duality
mapping  to define dual variables $\mu^\alpha_i$ as $\mu^{\alpha}_i=
U_D\sigma^{\alpha}_iU_D^\dagger$. Then the above relations lead to
$\mu_1^x=\sigma^z_N,\ \mu^x_i=\sigma^z_{r(i)} \sigma^z_{r(i)+1},\
i=2,\cdots,N$. On the other hand,
$\mu^z_i=U_D\sigma^{z}_iU_D^\dagger=U_D\sigma^{z}_i\sigma^z_{i+1}\times\cdots
\sigma^z_{N-1}\sigma^z_N\times\sigma^z_NU_D^\dagger$, so that, by the
duality mapping above,  reduces to
$\mu^z_i=\prod_{m=i}^N\sigma^x_{r(m)}=\prod_{m=1}^{N-1+i}\sigma^x_m$. 
Similarly, the Jordan-Wigner dictionary \cite{GJW} gives rise to a bond
algebra  mapping when applied to $d=1$ spin and spinless Fermi  systems.
The explicit exchange statistics transformation can be  derived by
solving for one set of bonds in terms of the other. It can be shown that
there is no  Jordan-Wigner transformation that  relates two {\em local
Hamiltonians} in dimensions $d>1$:  By examining the product of bonds
around closed loops an inconsistency  is found if local spin-less Fermi
bilinears could be mapped to local spin  terms and vice versa. In the
following we disregard boundary terms without further comments.

{\it Dualities and Self-dualities in Quantum Statistical Mechanics.} The
$d=3$ orbital compass (OC) model 
\begin{align}
H_{\sf OC}= -\sum_{\vec{\imath}} [&J_x S^{x}_{\vec{\imath}}
S^{x}_{\vec{\imath}+\vec{e}_1}+J_y S^{y}_{\vec{\imath}}
S^{y}_{\vec{\imath}+\vec{e}_2}+ J_zS^{z}_{\vec{\imath}}
S^{z}_{\vec{\imath}+\vec{e}_3}] \nonumber
\end{align}
($S^{\alpha}_{\vec{\imath}}=\frac{1}{2}\sigma^{\alpha}_{\vec{\imath}}$)   
has been proposed \cite{orbital_compass} to study orbital ordering in
transition metal compounds.
A still interesting yet simplified scenario for orbital ordering is
provided by the planar OC model (POC)
\begin{eqnarray}
H_{\sf POC}[J_x,J_y]=-\sum_{\vec{\imath}} (J_x
\sigma^x_{\vec{\imath}}\sigma^x_{\vec{\imath}+\vec{e}_1}+
J_y\sigma^y_{\vec{\imath}}\sigma^y_{\vec{\imath}+\vec{e}_2})
\end{eqnarray}
Its bond algebra ${\cal A}_{H_{\sf POC}}$ is generated by 
$\{\sigma^x_{\vec{\imath}}\sigma^x_{\vec{\imath}+\vec{e}_1},
\sigma^y_{\vec{\imath}} \sigma^y_{\vec{\imath}+\vec{e}_2}
\}_{\vec{\imath}}$, and it is specified by a few relations: Each bond
(i) squares to one, (ii) anti-commutes with the four other bonds which
share any of its vertices, and (iii) commutes with all other bonds. The
mapping $\Phi(\sigma^x_{\vec{\imath}}\sigma^x_{\vec{\imath}+\vec{e}_1})=
\sigma^y_{\vec{\imath}+\vec{e}_1}\sigma^y_{\vec{\imath}+\vec{e}_1+\vec{e}_2}$,
$\Phi(\sigma^y_{\vec{\imath}}\sigma^y_{\vec{\imath}+\vec{e}_2})=
\sigma^x_{\vec{\imath}+\vec{e}_2}\sigma^x_{\vec{\imath}+\vec{e}_2+\vec{e}_1}
$, preserves every relation among bonds, showing a self-duality under 
$J_x\leftrightarrow J_y$.
The POC Hamiltonian is dual as well \cite{NF} to the Xu-Moore (XM)
Hamiltonian \cite{XM}
\begin{equation}
H_{\sf XM}[j,h]= -\sum_{\vec{\imath}} (j
\square{\sigma^z}_{\vec{\imath}}+h \sigma^x_{\vec{\imath}}) , 
\label{XM}
\end{equation}
(with $\square {\sigma^z}_{\vec{\imath}} =\sigma^z_{\vec{\imath}} 
\sigma^z_{\vec{\imath}+\vec{e}_1} \sigma^z_{\vec{\imath}+\vec{e}_1
+\vec{e}_2} \sigma^z_{\vec{\imath}+\vec{e}_2}$) 
which was introduced as a simplified model for some aspects of quantum
phase transitions in $p+ip$ superconducting arrays. The duality comes
from the mapping of bonds
$\Phi(\sigma^x_{\vec{\imath}}\sigma^x_{\vec{\imath}+\vec{e}_1})=\square
{\sigma^z}_{\vec{\imath}},\ \Phi(\sigma^y_{\vec{\imath}}
\sigma^y_{\vec{\imath}+\vec{e}_2})=\sigma^x_{\vec{\imath}+\vec{e}_2}$,which
is indeed given by a unitary $U_D$. Thus $ U_DH_{\sf
POC}[J_x,J_y]U_D^{\dagger} =H_{\sf XM}[J_x,J_y]$, and these two models
must have the same phase diagram. In spite of this, the
quantum($d$)-to-classical($d+1$) mapping is much easier  for $H_{\sf
XM}$ than for $H_{\sf POC}$, another manifestation of the power of
duality transformations and a useful fact if one wants to perform, say,
quantum Monte Carlo simulations. The self-duality of the XM Hamiltonian
\cite{XM} can be deduced from the self-duality of the POC model and the
duality just described, or directly as an automorphism of its bond
algebra.  Applied to the elementary degrees of freedom
$\{\sigma^x_{\vec{\imath}},\sigma^z_{\vec{\imath}}\}$, the automorphism
returns the non-local dual operators of \cite{XM}.

{\it Classical from Quantum Dualities.} The standard
quantum($d$)-to-classical$(d+1)$ connection establishes an equivalence
between quantum (as unitary mappings) and classical dualities. Take for
example the XM Hamiltonian $H_{\sf XM}[j,h]$ of Eq. \eqref{XM}.  Its
classical rendition is $\left(\half \sinh(2J^*)\right)^{\Omega/2}{\cal
Z}[J,K]$, with  ${\cal Z}[J,K]\equiv
\sum_{[\sigma]}e^{\sum_{\vec{\imath},t}\left(J\square{\sigma^z}_{\vec{\imath}}(t)+
K\sigma_{\vec{\imath},t}\sigma_{\vec{\imath},t+1}\right)} $,
$J=j\frac{\Delta\tau}{N_t}$, $J^*=h\frac{\Delta\tau}{N_t}$, and
$K=-\half\ln\tanh\left(h\frac{\Delta\tau}{N_t}\right)$. The length along
the time axis $N_t \gg 1$. Similarly, $H_{\sf XM}[h,j]$ maps to
$\left(\half \sinh(2J)\right)^{\Omega/2}{\cal Z}[J^*,K^*]$, with
$K^*=-\half\ln\tanh\left(j\frac{\Delta\tau}{N_t}\right)$. It follows
already that $\sinh 2J\sinh 2K^*=1=\sinh 2J^*\sinh 2K$, yet nothing in
principle guarantees any relation between ${\cal Z}[J,K]$ and ${\cal
Z}[J^*,K^*]$ so far.  Now, due to the quantum self-duality $H_{\sf
XM}[j,h]=U_DH_{\sf XM}[h,j]U_D^{\dagger}$, we have that ${\sf Tr}
\exp\left(-\Delta\tau H_{\sf XM}[j,h]\right)= {\sf
Tr}\exp\left(-\Delta\tau H_{\sf XM}[h,j]\right)$. Hence  $\frac{{\cal
Z}[J,K]}{\left(\half \sinh(2J)\right)^{\Omega/2}}=\frac{{\cal
Z}[J^*,K^*]}{\left(\half \sinh(2J^*)\right)^{\Omega/2}}$, which is
indeed  the classical self-duality obtained in \cite{XM} by considerably
more laborious classical methods.

{\it Emergent Dualities.} A  (self-)duality  {\em can emerge} in a
sector of a theory (e.g., for particular subsets of couplings, or low
energy subspace). The projection of a bond algebra onto a sector of the
full Hilbert space generates a new bond algebra 
that may have (self-)dualities not present in the full model. An example
is provided by the Quantum Dimer Model (QDM) \cite{RK} defined on the
orthonormal set of dense dimer coverings of a lattice. The QDM
Hamiltonian reads
\begin{eqnarray}
H_{\sf QDM}&=&\sum_{\Box} \left[
-t\left(\left|\vv\right\rangle\left\langle\hh\right| +
\left|\hh\right\rangle\left\langle\vv \right|\right)\right.
\nonumber\\
&+&\left.v\left(\left|\vv\right\rangle\left\langle\vv\right|+
\left|\hh\right\rangle\left\langle\hh
\right|\right) \right],
\label{QDM}
\end{eqnarray}
with the sum performed over all elementary plaquettes. The QDM contains
both a kinetic ($t$) term that flips one dimer tiling of any plaquette
to another (a horizontal covering to a vertical one and vice versa), and
a potential ($v$) term. At the (so-called) RK point $t=v$ \cite{RK}, the
ground states are equal amplitude superpositions of dimer coverings. If
$P_{g}$ is the projection operator onto the ground state sector, then
$P_{g} [\left(\left|\vv\right\rangle\left\langle\hh\right| +
\left|\hh\right\rangle\left\langle\vv \right|\right)]_{\Box} P_{g}
=P_{g}
[\left(\left|\vv\right\rangle\left\langle\vv\right|+\left|\hh\right\rangle\left\langle\hh
\right|\right)]_{\Box} P_{g} = x_{\Box} P_{g}$, with $x_{\Box} = 0$ or
$1$ on the particular plaquette $\Box$ where $
[\left(\left|\vv\right\rangle\left\langle\hh\right| +
\left|\hh\right\rangle\left\langle\vv \right|\right)]_{\Box} $ flips the
dimer in the plaquette $\Box$. At the  RK point, the projected
Hamiltonian becomes $P_{g} H_{\sf QDM} P_{g} =0$. Since both the kinetic
($t$) and potential ($v$) terms are given by $x_{\Box} P_{g}$ within the
ground state sector, the kinetic and potential operators can be
interchanged without affecting the bond algebra. This self-duality
emerges exclusively in the ground state sector of the QDM at the RK
point.

{\it Dualities in Quantum Field Theory (QFT).} An elementary application
of our technique is provided by a free massless scalar field in $1+1$
dimensions \cite{witten}, with Hamiltonian  $H= \frac{1}{2} \int dx\
[\pi^{2}(x,t)+ \left (\frac{\partial\phi(x,t)}{\partial x}\right
)^{2}],$ and $[\phi(x,t),\pi(x',t)]=i\delta(x-x')$. (With obvious
modifications, this Hamiltonian describes a taut string.) To study this
model's bond algebra, it is convenient to discretize it, with lattice 
spacing $a$, i.e. 
$\frac{a}{2} \sum_i [\pi_i^2+ (\phi_{i+1}-\phi_i)^2/a^2]$. The
automorphism  $\pi_i \mapsto -(\phi_{i+1}-\phi_i)/a,\ \
(\phi_{i+1}-\phi_i)/a \mapsto -\pi_{i+1}$ preserves the canonical
commutation relations. The dual variables provide a convenient way to
study this self-duality in the continuum. Their discrete form is
$\tilde{\phi}_i=a\sum_{m=i+1}^\infty \pi_m, \ \
\tilde{\pi}_i=-(\phi_{i+1}-\phi_{i})/a$. Now we can let $a$ go to zero
to obtain dual variables in the continuum:
$\tilde{\pi}(x,t)=U_D\pi(x,t)U_D^{\dagger}=-\frac{\partial
\phi}{\partial x}(x,t)$,
$\tilde{\phi}(x,t)=U_D\phi(x,t)U_D^{\dagger}=\int_x^\infty dy\
\pi(y,t)$. These are toy examples of solitonic variables. In general,
self-dualities can be destroyed by coupling the system to sources, but
this is not necessarily the case. Consider the scalar field now coupled
to external classical sources $A,E$: $H^{A,E}=\int dx\left [ 
\frac{1}{2} (\pi-\lambda A)^2+\frac{1}{2}\Bigl (\frac{\partial
\phi}{\partial x}\Bigr )^2-\lambda E \phi \right ]$.
The self-duality maps $H^{A,E}$ to
\begin{align}
H&^{\tilde{A},\tilde{E}}\!\!=\!\!\int \!\! dx \left [
\frac{1}{2}(\pi-\lambda \tilde{A})^2\!\!+\frac{1}{2}
\Bigl (\frac{\partial \phi}{\partial x}\Bigr )^2\!\!-\lambda
\tilde{E}\phi+\mbox{c-number} \right ] \!\!.\nonumber
\end{align}
The self-duality survives this  coupling to external sources, with dual
sources $\tilde{A}(x,t)=\int_{-\infty}^{x} dw\ E(w,t),\ \
\tilde{E}(x,t)= \frac{\partial A}{\partial x}(x,t)$.

Next we consider ${\mathbb Z}_N$ gauge field theories (GFTs) defined on
a Euclidean $3+1$-dimensional lattice. The interest in these theories
grew out of 't Hooft studies on quark  (charge) confinement in pure
$SU(N)$ gauge theories \cite{thooft}, that suggest that their most
important degrees of freedom  near a confinement-deconfinement phase
transition are the field configurations taking values in the center
subgroup of $SU(N)$, ${\mathbb Z}_N$. To explore this scenario, several
author  considered Wilson's action for Euclidean lattice GFTs
\cite{yoneya}, $S=-\frac{1}{g^2}(\sum_{\Box} \mbox{Re Tr}(U_
{i,j}U_{j,k}U_{k,l}U_{l,i}-1))$, restricting the fields to take values
in ${\mathbb Z}_N$. This is the model we are going to study, thus  
$U_{i,j}$ stands for a $N$th root of unity attached to the oriented link
with  endpoints $i,j$, and $U_{i,j}=U^{*}_{j,i}$. 
In the axial gauge the action simplifies 
\begin{eqnarray}
S=-\frac{1}{2g^2}\sum_n
\sum_{i=1}^3 \left [ \cos(\theta^i_{n+e_4}-\theta^i_n)+\cos(\Theta^i_n)
\right ] ,\nonumber
\end{eqnarray}
($\Theta^1_n=\theta^3_{n+e_2}-\theta^3_n-\theta^2_{n+e_3}+\theta^2_n$),
and cyclic permutations thereoff. The goal is to learn about duality
properties of amplitudes in QFTs, as  given by a path integral over
field configurations. Computation of a vacuum to vacuum amplitude
$\langle 0 \, {\sf out}\vert 0 \, {\sf in}\rangle$ amounts to evaluating
a partition function.  Thus we  can apply the bond algebra technique to
look for self-dualities in QFTs that are more conveniently  quantized
through path integrals. To proceed, we need to compute the quantum
Hamiltonian equivalent to  the gauge fixed action given above. This is a
difficult task for arbitrary $N$, but the computations were done  (in a
different context) in \cite{ortiz}. Using these (the coupling $K$
depends on $N$ and $g$ \cite{K}) 
$$H= - \sum_n\sum_{i=1}^3  \Big[K
V_n^i + \frac{1}{4g^2}\Delta\theta^{i }_n \Big]+h.c.,$$
where
$\Delta\theta^{3}_n=U^1_nU^2_{n+e_1}U^{1\dagger}_{n+e_2}U^{2\dagger}_n$,
and cyclic permutations. There are now $N\times N$ unitary matrices $U,\
V$ on each link $(n,e_i),\ i=1,2,3$ of a cubic lattice, ($U^i_n,\ V^i_n$
denote matrices on the link $(n,e_i)$). The $U$s and $V$s satisy $
(U^{i}_n)^N=(V^i_n)^N=1$, $V^i_nU^{i}_n=\omega U^{i}_nV^i_n$
($\omega=e^\frac{2\pi i}{N}$), i.e., Weyl's group relations, and
matrices on different links commute. $\mathbb Z_N$ GFTs have been known
for many years to be self-dual for $N=2,3,4$, and it was conjectured
that they are no longer self-dual for $N\geq 5$ \cite{yoneya}. {\it We
can prove that these theories remain self-dual for all $N$}, as the
mapping of bonds
\begin{align}
V_n^1\mapsto \Delta\theta^1_n&,\ \ \
\Delta\theta^1_n \mapsto V^{1\ \dagger}_{n-e_1+e_2+e_3}\nonumber
\\
V_n^2 \mapsto \Delta\theta^2_{n-e_1+e_2}&,\ \ \
\Delta\theta^2_n \mapsto V^{2\ \dagger}_{n+e_3}\label{gaugeduality} \\
V_n^3\mapsto \Delta\theta^3_{n-e_1+e_3}&,\ \ \
\Delta\theta^3_n \mapsto V^{3\ \dagger}_{n+e_2}\nonumber
\end{align}
shows. $U_D^2$ is a new discrete symmetry of this problem, but $U_D^4=1$
up to a lattice translation. For large $N$, these gauge theories are
known to display three phases, two of them connected through a 
confinement-deconfinement phase transition \cite{jersak}. The
self-duality fixes the self-dual coupling $g^*$ at $4g^{*2}K^*=1$
\cite{K}, {\it which gives the exact self-dual coupling for every $N$}
(so far only known analytically for $N=2,3,4$). On the other hand,  it
is shown in \cite{ortiz} (using our approach) that the isotropic
$d+1=1+1$ $N$-state vector Potts model has a self-dual point at coupling
$J^*$ given by precisely an equivalent relation $2K^*=J^*$. 
{\it Thus our results  explain the puzzling fact} \cite{yoneya} that the
isotropic classical $d+1=1+1$  $N$-state vector Potts model and the
$d+1=3+1$ $\mathbb Z_N$ GFT share identical  self-dual relation: first,
both bond algebras (though non-isomorphic) are based on  the Weyl
algebra, and  admit self-duality mappings; and second, both models have
quantum couplings satisfying the equation in \cite{K}. 
The compactness of degrees of freedom  (i.e., angular variables), is
required for a phase transitions to occur.
On one hand, Polyakov \cite{polyakov} showed  that compact QED displays
no phase transitions in $2+1$ dimensions. On the other, we can show that
in the limits $N\rightarrow \infty$, $a\rightarrow 0$,
\eqref{gaugeduality} reduces to the well know self-duality of vacuum QED
in $3+1$ dimensions $E\mapsto B,\ B\mapsto -E$, which has no phase
transitions.  We argue that since the self-duality emerges only in $3+1$
dimensions, it is important in  triggering the phase transitions of
these GFTs. So, the presence of both compactness and  self-duality are
crucial for the existence of a confinement-deconfinement phase
transition. 

In summary, we developed a unifying and systematic framework for
dualities, providing a new perspective to  unveil them:
(self-)dualities  (exact or emergent) can be investigated as {\em
homomorphisms of bond algebras}. The power  of this algebraic approach
was exploited to obtain new self-dualities of confining Abelian  GFTs in
$3+1$ dimensions, a new discrete symmetry of these theories, and their
self-dual couplings analytically. We prove that the puzzling connections
between these GFTs and some confining theories in $1+1$ dimensions (vector
Potts model) result from these two models having similar algebraic
structures and self-dualities. 
Self-dualities are more easily discovered as automorphisms of bond
algebras (quantum) than as relations between partition functions
(classical). Furthermore, they can generate otherwise hidden symmetries.
Known classical dualities derived in the literature by Fourier
transformation \cite{Wu} can be obtained by our technique. Thus this
work hints at a deep connection between operator algebra homomorphisms 
and the Fourier transform to be at the root of the equivalence between
classical and quantum dualities. 
Our approach to (self-)dualities is applicable to any  system, and
clears the way for the development of approximation schemes that
preserve these peculiar symmetries.

E.C. thanks V. Lunts for helpful discussions.

{\bf Unified approach to Quantum and Classical Dualities: Supplementary Material}

In this section, we present further applications of our method.  Our
approach provides a new perspective on  all dualities in physics. With
few exceptions, the study of (self-)dualities in classical  and quantum
systems has been  dominated \cite{SSavit} by techniques that amounts to
a change of (classical) variables in partition functions. Our novel
operator technique is different enough to require careful examination
through several examples.

{\it Dualities of the extended Toric Code (TC) Model.} The TC model
\cite{kitaev} was recently extended to include an external magnetic
field $\vec{h}$ \cite{prokofev}
\begin{eqnarray}
\!\!H_{\sf ETC}=  \!-J_{x} \! \sum_{s} \!A_{s} - \! J_{z}\!
\sum_{p} \! B_{p} - \!\! \sum_{\langle ij \rangle} ( h_x
\sigma^x_{ij}+ h_z\sigma^z_{ij} ). \label{ETCEQ}
\end{eqnarray}
In Eq. (\ref{ETCEQ}),  on each {\it link} of a square lattice there is
an spin-1/2 operator, $A_{s} = \prod_{j} \sigma^{x}_{js}$ represents a
product over the $x$-component of the four spins that have $s$ as a
common vertex, and $B_{p}= \prod_{j} \sigma^{z}_{jp}$ is the product of
the four $\sigma^{z}_{jp}$ operators that belong to the plaquette
($\Box$) $p$. The quantum phase diagram of $H_{\sf ETC}$ has  been
studied quite recently \cite{ prokofev,vidal}. It shows reflection
symmetry relative to the line $h_x=h_z$, and a multi-critical  point on
this line as well \cite{prokofev, vidal}. The special role of the
condition $h_x=h_z$  can be understood in terms of a quantum
self-duality. Drawing straight lines through the centers of the bonds,
$H_{\sf ETC}$ can we re-written in terms of spins lying on the vertices
of another square at a $45$ degree angle with the original one. On this
lattice,  the mapping
$\Phi(\sigma^x_{\vec{\imath}})=\sigma^z_{\vec{\imath}+\vec{e}_2}$,
$\Phi(\sigma^z_{\vec{\imath}})=\sigma^x_{\vec{\imath}+\vec{e}_2}$
extends to a self-duality automorphism of ${\cal A}_{H_{\sf ETC}}$ that
exchanges $J_x$ with $J_z$ and simultaneously $h_x$ and $h_z$. The
reflection symmetry $J_{x} \leftrightarrow J_{z}$ and $h_{x}
\leftrightarrow h_{z}$ relates to  the Wegner duality \cite{wegner,
Skogut} and the more general self-duality of an Ising matter coupled
gauge theory. On a cubic lattice, the action for the latter reads
\cite{fradkin}
\begin{eqnarray}
S =  -J \sum_{\langle i j \rangle}  \sigma_{i} \eta_{ij}
\sigma_{j} - K \sum_{\Box} \Box \eta , \label{z2}
\end{eqnarray}
where matter fields $\{\sigma_{i}\}$ live at lattice sites $i$, while
gauge fields $\eta_{ij}$ live on the links connecting sites $i$ and $j$.
This action is self-dual under
\begin{eqnarray}
J \to \tilde{J} = K, ~~~ K \to \tilde{K} = - \frac{1}{2} \ln \tanh
J, \label{weg}
\end{eqnarray}
which can be derived as a consequence of the self-duality of the ETC
Model. This follows from the Euclidean representation of Eq.
(\ref{ETCEQ}) given by the 3 (or 2+1)-dimensional action of Eq.
(\ref{z2}) \cite{prokofev}, with
\begin{eqnarray}
J = h_{z} \Delta \tau, ~~~K = -\ln \tanh (h_{x} \Delta \tau) ,
\label{jkd}
\end{eqnarray}
where $\Delta \tau$ is the discretization step in the imaginary time
direction. Inserting the duality $\sigma_{ij}^{x} \leftrightarrow
\sigma_{ij}^{z}$  into Eq. (\ref{jkd}), we derive the gauge theory
dualities of Eq. (\ref{weg}). The {\em Wegner duality} between the Ising
model ($K=0$) and the gauge theory ($J=0$) is a particular case of Eq.
(\ref{jkd}). The Ising matter coupled gauge system offers  another
example of an emergent duality. Take the system of Eq. (\ref{z2}) in the
limit $K \to \infty$, that enforces a projection onto a space in which
$\eta_{ij} = w_{i} w_{j}$ ($w_{i} = \pm 1$). Setting $s_{i} \equiv
\sigma_{i} w_{i}$, we obtain  a general $d$-dimensional rendition of the
matter coupled gauge theory of Eq. (\ref{z2}), the $d$-dimensional Ising
model. The KW self-duality of the classical $d=2$ Ising model ($J \to
\tilde{J} = \tanh^{-1}(\exp[-2J])$) appears as an emergent duality in
the $K \to \infty$ limit of the system of Eq. (\ref{z2}). In this limit,
the bonds $z_{ij} = \sigma_{i} \eta_{ij} \sigma_{j}$ \cite{nprd} satisfy
the constraints ${\cal{C}}_{\Box} $: $ \prod_{ij  \in \Box} z_{ij} =1$
for all plaquettes $\Box$. In the projected subspace in which the
constraints $\{{\cal{C}}_{\Box}\}$ are satisfied, the algebras of the
matter coupled gauge theory of Eq. (\ref{z2}) and the Ising model are
identical. For finite $K$, the Ising matter coupled gauge system of Eq.
(\ref{z2}) \cite{fradkin}, is dual in $d=2$ to an Ising model in a
uniform magnetic field \cite{nprd} which does not obey the KW relations.

Next, we introduce a model that is dual to the ETC model for arbitrary
couplings. The dual Hamiltonian is
\begin{eqnarray}
H &=& - J_{x} \sum_{\Box} \Box \eta^{x}- J_{z} \sum_{i}
\mu_{i}^{z}
\nonumber \\
&&- h_{x} \sum_{\langle ij \rangle} \mu_{i}^{x} \mu_{j}^{x}
\eta_{ij}^{x} - h_{z} \sum_{\langle ij \rangle} \mu_{i}^{z}
\mu_{j}^{z} \eta_{ij}^{z} . \label{ETC'}
\end{eqnarray}
In the particular case $J_{z}= h_{z} =0$, this is the classical Ising
matter coupled lattice gauge theory.  For a square lattice of $N$ sites
with periodic boundary conditions, there are $N$ matter fields of the
type $\{\mu_{i}^\alpha\}$ and $(2N)$ gauge fields
$\{\eta_{ij}^\alpha\}$. For a dual system on  a square lattice of $N'$
sites, there are $(2N')$ spin fields in the Hamiltonian of Eq.
(\ref{ETCEQ}). As the bond algebras in the two systems defined by the
Hamiltonians of Eqs. (\ref{ETCEQ}), and (\ref{ETC'}) are the same then,
ignoring additional global constraints stemming from boundary
conditions, in computing the partition function ${\cal Z}$, there is a
one-to-one correspondence between the terms for the two dual models. If
the only element with a non-vanishing trace is the  identity operator
then, when $N' = (3N)/2$ then each of the terms in the expansion of
${\cal Z}$ will give an identical contribution.

Finally, if
$h_{x} =0$, a model dual to Eq. (\ref{ETCEQ}) is given by
\begin{eqnarray}
H = - J_{x} \sum_{i} \mu_{i}^{z} - J_{z} \sum_{i} \Box \mu^{z} -
h_{z} \sum_{\langle i j \rangle} \mu_{i}^{x} \mu_{j}^{x} .
\label{x0}
\end{eqnarray}
The Hamiltonian of Eq. (\ref{x0}) is determined by the $N'$ variables
$\{\mu^\alpha_{i}\}$. This duality  is exact if we scale the number of
variables accordingly: $N' = 2N$.

{\it Beyond $S$=$\frac{1}{2}$. }   The Blume-Emery-Griffiths (BEG) model
\cite{BEG} -in its square lattice version (coordination $z$=4)- is given
by an $S$=1 Ising-like Hamiltonian
\begin{eqnarray}
\beta H_{\sf BEG}=-\sum_{\langle ij \rangle} (J S^z_iS^z_j + K
(S^z_i S^z_j)^2) + D \sum_i (S^z_i)^2.
\end{eqnarray}
The BEG model was developed in the context of liquid $^3$He--$^4$He
mixtures. At particular values for the parameters $K=3J$, and $D=2zJ$,
this model becomes  the isotropic $q=3$ Potts model, $\beta H_{\sf
P}=-2J \sum_{\langle ij \rangle} \delta(S^z_i, S^z_j)$, ($S^z=\pm 1,
0$).

It  can be shown, using the transfer matrix technique, that the
$d$=1 quantum BEG model is equivalent to
\begin{equation}
H_{\sf QBEG}=-B\sum_i\sqrt{3}v_i-J\sum_i\frac{2}{\sqrt{3}}\omega_i
\end{equation}
where the bonds are
\begin{eqnarray}
\sqrt{3}v_i \!\!&=&\!\! 1+\sqrt{2}
S^x_i+\frac{D_{xy}}{2B}((S^{+}_i)^2+(S^{-}_i)^2)+\frac{D_\perp}{B}
(S^z_i)^2 \nonumber \\
\frac{2}{\sqrt{3}} \omega_i  \!\!&=& \!\!2+ \!S^z_i S^z_{i+1}
\!+\! \frac{K}{J} (S^z_i S^z_{i+1})^2  \! - \!\frac{D}{4J}
((S^z_i)^2+(S^z_{i+1})^2) \nonumber
\end{eqnarray}
with dual coupling constants  $ B=\frac{\Lambda_3}{2+\Lambda_3^2} \ln
\lambda_2+ \frac{\Lambda_2}{2+\Lambda_2^2} \ln \lambda_3\nonumber$,
$D_{xy}=-\frac{\ln\lambda_1}{2}+\frac{1}{2+\Lambda_3^2} \ln \lambda_2+
\frac{1}{2+\Lambda_2^2} \ln \lambda_3$, and
$D_\perp=\frac{\ln\lambda_1}{2}+ \frac{1-\Lambda_3^2}{2+\Lambda_3^2}\ln
\lambda_2+ \frac{1-\Lambda_2^2}{2+\Lambda_2^2} \ln \lambda_3$. The
parameters $\lambda_{1,2,3}$ are given by  $\lambda_1=2
e^{(K-\frac{D}{2})} \sinh J$, $\lambda_2+\lambda_3=2 e^{(K-\frac{D}{2})}
\cosh J+1$, and  $\lambda_2 \lambda_3=2 (e^{(K-\frac{D}{2})} \cosh J -
e^{-\frac{D}{2}})$. Finally, $\Lambda_\alpha=e^{\frac{D}{4}}
(1-\lambda_\alpha)$.

The classical BEG system has a tricritical point, which  is mapped to a
quantum critical point of $H_{\sf QBEG}$. At $K=3J$, $D=8J$ (the Potts
limit \cite{mittag}), the bonds $\omega_i$ and $v_i$ satisfy the Hecke
relations $\omega_i v_{i+1} \omega_i - \omega_i =
v_{i+1}\omega_iv_{i+1}-v_{i+1} \ , \ \omega_i \omega_i = \sqrt{3}\,
\omega_i$,  $ v_i \omega_{i} v_i - v_i =
\omega_{i}v_i\omega_{i}-\omega_{i} \ , \ v_i v_i = \sqrt{3} \, v_i$.
Therefore, a quantum self-duality defined on  generators by $v_i
\rightarrow \omega_i$, $\omega_i \rightarrow v_i$, fixes the self-dual
point  at $3B=2J$.  Thus by algebraic means alone, we can determine the
tricritical temperature $T_c$ from the dual coupling $B$:
$k_BT_c=\frac{2J}{\ln(1+\sqrt{3})}$, in agreement with the result
obtained through other methods \cite{baxter}.

{\it Self-duality of vacuum Quantum Electrodynamics (QED).} 
\begin{figure}[t]
\begin{center}
\includegraphics[width=0.75\columnwidth]{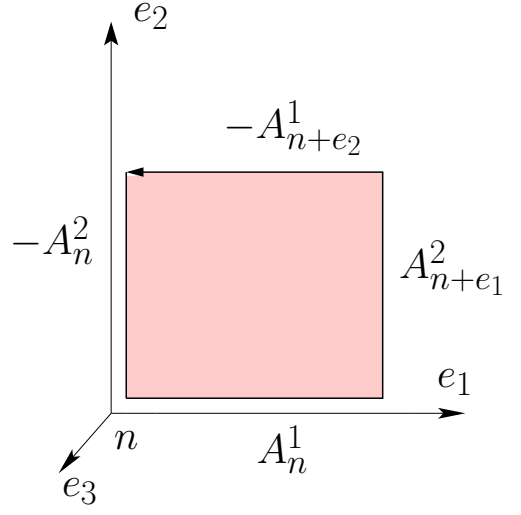}
\end{center}
\caption{The plaquette variable
$\Delta\theta_n^3=A^2_{n+e_1}-A^2_n-A^1_{n+e_2}+A^1_{n}$.}
\label{plaquette}
\end{figure}
The classical  (self-)duality of electromagnetic fields in the absence
of sources (in vacuo) $E\rightarrow B,\ \ B\rightarrow -E$ is one of the
oldest examples of a self-duality available. The bond algebra formalism
gives insight into how this self-duality extends to non-compact vacuum
QED. To see the connection with the self-duality we found for
$\mathbb{Z}_N$ lattice gauge theories, we quantize electromagnetism in
the axial gauge, $\phi \equiv 0$. In this gauge, the classical fields 
in terms of the vector potential $A$ are $E=-\frac{\partial A}{\partial
t}, \ \ \ B=\nabla\times A$. If we next apply canonical quantization to
electromagnetism written in this gauge, we get vacuum QED in the form 
$$H=\int d^3x\ \frac{1}{2} (\Pi^2+(\nabla \times A)^2),$$  
together with canonical commutation relations
$[A_m(x,t),\Pi_l(y,t)]=i\delta_{m l}\delta(x-y)$, where we made the
canonical substitution $\frac{\partial A}{\partial t}=-E \rightarrow
\Pi$. Furthermore, from gauge invariance, we have that the subspace of
physical states (gauge invariant states) of the full Hilbert space is
specified by the so called Gauss constraint: $\nabla \cdot \Pi\
\vert\mbox{physical}\rangle=0$ (physically, this means we only keep
those states on which $\nabla\cdot E=0$). To investigate the bond
algebra of this Hamiltonian, it is convenient to discretize the theory
and consider it on a $d=3$ cubic lattice of lattice spacing 
$a$. The Hamiltonian then reads ($n=(n^1,n^2,n^3)$)  
$$H=\sum_n a^3\sum_{i=1}^3\frac{1}{2}(\Pi_n^i)^2+\frac{1}{2}(\Delta\theta_n^i)^2,$$ 
where $[A_n^i,\Pi_m^k]=i\delta^{i k}\delta_{n,m}$, and
\begin{eqnarray}
\Delta\theta_n^1=(A^3_{n+e_2}-A^3_n-A^2_{n+e_3}+A^2_{n})/a \nonumber \\
\Delta\theta_n^2=(A^1_{n+e_3}-A^1_n-A^3_{n+e_1}+A^3_{n})/a \nonumber \\
\Delta\theta_n^3=(A^2_{n+e_1}-A^2_n-A^1_{n+e_2}+A^1_{n})/a \nonumber
\end{eqnarray}
are the discretized form of the components of $\nabla\times A$. We call
the interaction terms $\Delta\theta$ plaquette interactions, borrowing
the terminology we used  with $\mathbb{Z}_N$ gauge theories (see  Fig.
1).
\begin{figure}[h]
\begin{center}
\includegraphics[width=.90\columnwidth]{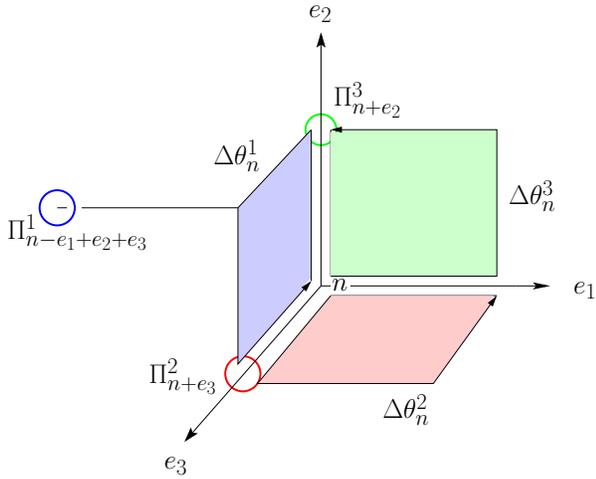}
\end{center}
\caption{Schematic of the bond algebra mapping \eqref{dual_mapping}.
Associated with each lattice site $n$, there are three $\Pi$ fields and
three $\Delta \theta$ plaquettes. 
The $\Delta \theta$ plaquettes at site $n$ map by
\eqref{dual_mapping} to displaced $\Pi$s, each colored plaquette mapping
to the $-\Pi$ at the correspondingly colored site.}
\label{nicecube}
\end{figure}
The first thing to notice is that only four plaquette variables
$\Delta\theta$ have non-trivial commutation relations with any given
momentum component $\Pi^i_n$ at site $n$ and similarly, only four $\Pi$
fields have non-trivial commutation relations with any given
$\Delta\theta^i_n$. Thus we have a good case for an automorphism that
exchanges $\Pi \leftrightarrow \Delta\theta$. In fact, one such
automorphism is given by
\begin{align}
\Pi^1_n\mapsto\tilde{\Pi}^1_n=\Delta\theta^1_n,\ \ 
\Delta\theta^1_n\mapsto\tilde{\Delta}\theta^1_n=-\Pi^1_{n-e_1+e_2+e_3}\nonumber\\
\Pi^2_n\mapsto\tilde{\Pi}^2_n=\Delta\theta^2_{n-e_1+e_2},\ \ 
\Delta\theta^2_n\mapsto\tilde{\Delta}\theta^2_n=-\Pi^2_{n+e_3}\label{dual_mapping} \\
\Pi^3_n\mapsto\tilde{\Pi}^3_n=\Delta\theta^3_{n-e_1+e_3},\ \
\Delta\theta^3_n\mapsto\tilde{\Delta}\theta^3_n=-\Pi^3_{n+e_2}.\nonumber
\end{align}
The geometry of this mapping is clarified in the Figs. 2 and 3. A
moment's reflection makes it clear that this mapping is  nothing other
than  $E\mapsto B,\ B\mapsto -E$, the quantum descendant of the
classical electromagnetic self-duality. As explained in the main body of
our paper, the self-duality mapping serves the double purpose of
establishing the existence of a self-duality and defining the dual
variables. We have explicitly shown the dual variables above, naming
them $\tilde{\Pi}$ and $\tilde{A}$. The dual vector potential is defined
implicitly by the above relations, as, for instance,
$\tilde{\Delta}\theta_n^1=(\tilde{A}^3_{n+e_2}-\tilde{A}^3_n-
\tilde{A}^2_{n+e_3}+\tilde{A}^2_{n})/a$.

It is interesting to compute explicitly the dual variables in the
continuum limit $a\rightarrow 0$, paralleling the discussion of dual
variables for the scalar field given in the main body of this paper. In
this limit, \eqref{dual_mapping} implies the relations
\begin{align}
\tilde{\Pi}(x,t)=\nabla\times A(x,t)\nonumber\\
\nabla\times \tilde{A}(x,t)=-\Pi(x,t)\nonumber
\end{align}
between the initial and the dual operator variables. Thus $\tilde{\Pi}$
is already explicitly given in terms of $A$. We need to solve the second
relation for $\tilde{A}$. It is not difficult to check that, {\it on
physical states} (on which $\nabla\cdot \Pi$ vanishes),
$$\tilde{A}(x,t)=-\frac{1}{4\pi}\nabla\times\int d^3w\ \frac{\Pi
(w,t)}{\vert x-w\vert}.$$

\begin{figure}[ht]
\begin{center}
\includegraphics[width=0.75\columnwidth]{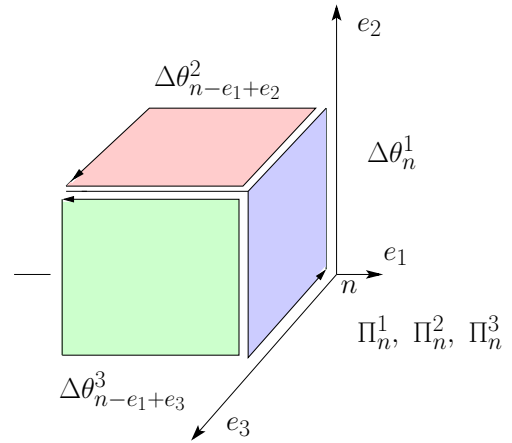}
\end{center}
\caption{The effect of the exchange duality of  \eqref{dual_mapping}
on the three 
$\Pi$ fields at site $n$.}
\label{nicecube2}
\end{figure}


\begin{thebibliography}{99}
\bibitem{irrelevant}
We call an operator ``irrelevant in the thermodynamic
limit" when its expectation value in any state, divided by the
system's size, vanishes in this limit.
\bibitem{K}
$K$ is computed from
$\frac{1}{2} \frac{\partial F_N(K)}{\partial
K}=e^{-\frac{1}{2g^2}(1-\cos\frac{2\pi}{N})}$, where
$F_N(K)=\sum_{n=0}^{N-1}e^{2K\cos(\frac{2\pi}{N}n)}$; $K^*$
is $K$ at $g=g^*$ \cite{ortiz}.
\end{thebibliography}
\end{document}